\documentclass[prd,aps,twocolumn,superscriptaddress]{revtex4-2}

\usepackage{amsmath}
\usepackage{cases}
\usepackage{graphicx}
\usepackage{natbib}
\usepackage{color}
\usepackage{xcolor}
\usepackage{tabularx}
\usepackage{bm}
\usepackage{threeparttable}
\usepackage{scrextend}
\usepackage{gensymb}
\usepackage{listings}
\usepackage{aas_macros}
\usepackage{orcidlink}

\begin{document}

\title{{Exploring Composition Mixing in Kilonova Ejecta with Ray-by-ray Simulations}}

\author{Ruocheng Zhai\textsuperscript{\orcidlink{0009-0004-1650-3494}}}
\email{rzhai@psu.edu}
\affiliation{Institute for Gravitation and the Cosmos, The Pennsylvania State University, University Park, PA 16802, USA}
\affiliation{Department of Astronomy \& Astrophysics, The Pennsylvania State University, University Park, PA 16802, USA}

\author{David Radice\textsuperscript{\orcidlink{0000-0001-6982-1008}}}
\affiliation{Institute for Gravitation and the Cosmos, The Pennsylvania State University, University Park, PA 16802, USA}
\affiliation{Department of Physics, The Pennsylvania State University, University Park, University Park, PA 16802, USA}
\affiliation{Department of Astronomy \& Astrophysics, The Pennsylvania State University, University Park, PA 16802, USA}

\author{Fabio Magistrelli\textsuperscript{\orcidlink{0009-0005-0976-7851}}}
\affiliation{Theoretisch-Physikalisches Institut, Friedrich-Schiller-Universit{\"a}t Jena, 07743, Jena, Germany}

\author{Sebastiano Bernuzzi\textsuperscript{\orcidlink{0000-0002-2334-0935}}}
\affiliation{Theoretisch-Physikalisches Institut, Friedrich-Schiller-Universit{\"a}t Jena, 07743, Jena, Germany}

\author{Albino Perego\textsuperscript{\orcidlink{0000-0002-0936-8237}}}
\affiliation{Dipartimento di Fisica, Università di Trento, via Sommarive 14, 38123 Trento, Italy}
\affiliation{INFN-TIFPA, Trento Institute for Fundamental Physics and Applications, via Sommarive 14, 38123 Trento, Italy}

\date{\today}

\begin{abstract}
    Binary neutron star merger (BNSM) ejecta are considered a primary repository of $r$-process nucleosynthesis and a source of the observed heavy-element abundances. We implement composition mixing into ray-by-ray radiation-hydrodynamic simulations of BNSM ejecta, coupled with an online nuclear network (NN). We model mixing via a gradient-based mixing approximation that evolves simultaneously with the hydrodynamics. We find that mixing occurs in regions where the electron fraction changes rapidly. While mixing smooths composition gradients in transition regions, it has a negligible impact on the heavy-element yields. This is because the primary $r$-process site (the equatorial ejecta) is initially homogeneous in free neutrons, leaving no strong gradients for mixing to act upon. In each angular ray, the abundances of the most produced elements are robust under mixing, while the less abundant ones are more affected. The total global abundances change only slightly from mixing, since each angular ray contributes its most abundant elements. Furthermore, the predicted kilonova light curves show only minor reddening, with differences below the detectability of state-of-the-art telescopes. In general, we do not observe significant effects from mixing in the time span of the $r$-process. Consequently, mixing only leads to minor variations in abundances and light curves in ray-by-ray simulations.
\end{abstract}

\maketitle

\section{Introduction}

Rapid neutron capture ($r$-process) nucleosynthesis is a primary mechanism producing heavy elements \citep{Cowan1991, Arnould2007}. Neutron-rich ejecta of binary neutron star mergers (BNSMs) are one of the leading candidates for explaining the current $r$-process element abundances \citep{Lattimer1974, Symbalisty1982, Eichler1989, Thielemann2017, Cowan2021, Perego2021, Arcones2023, Chen2024}. The first BNSM event confirmed via gravitational waves (GW), GW170817 \citep{Abbott2017a, Abbott2017b}, was accompanied by electromagnetic (EM) counterparts: the short gamma-ray burst GRB 170817A \citep{Goldstein2017} and the kilonova AT 2017gfo \citep{Coulter2017, Cowperthwaite2017}. This discovery has stimulated intense multi-messenger studies of BNSM outflows and their role in cosmic nucleosynthesis.

Numerical relativity (NR) simulations play an essential role in understanding the ejecta of BNSMs. They robustly predict that BNSM merger ejecta are composed of two components: dynamical and disk ejecta. The dynamical ejecta are generated on dynamical timescales. They consist of a component ejected by tidal torques \citep{Rosswog1999, Rezzolla2010, Radice2016, Dietrich2017, Bovard2017, Bernuzzi2020, Nedora2021}, and a component launched by shocks during and after the merger \citep{Hotokezaka2013, Bauswein2013, Sekiguchi2016, Dietrich2017, Bovard2017, Radice2018}. The disk ejecta is produced by the winds from mergers on a longer timescale \citep{Siegel2017, Siegel2018, Fujibayashi2018, Fernandez2019, Shibata2019, Nedora2019, Kiuchi2023}. The tidal ejecta are cold and neutron-rich, with electron fractions $Y_e \sim 0.1$ ($Y_e$ is the ratio between the net number of electrons and positrons, and baryons) \citep{Rosswog1999, Goriely2011}. The shock ejecta are heated by shocks and reprocessed by neutrino irradiation to higher $Y_e$ \citep{Wanajo2014, Radice2016}, especially in the polar regions with higher neutrino flux. A low-$Y_e$ ($\lesssim 0.25$) equatorial component, often found in the dynamical ejecta and late disk ejecta, produces lanthanides and actinides by $r$-process, leading to high opacity \citep{Kasen2013}. It powers kilonovae whose light curves peak in the infrared with timescales of days (red kilonovae; \citep{Metzger2014, Perego2017}). In the polar dynamical ejecta and most disk ejecta, higher $Y_e$ values suppress the synthesis of lanthanides and actinides. Consequently, this material retains low opacity, powering kilonovae that peak in the UV/optical with timescales of hours (blue kilonovae; \citep{Metzger2014, Perego2014, Martin2015, Metzger2019, Tanaka2017, Curtis2024}).

The evolution of the ejecta during the kilonova phase is challenging to model in the same way as BNSMs, given the orders-of-magnitude difference between the merger and kilonova timescales and lengthscales. Traditional models of kilonova evolution often assume homologously expanding ejecta without radiation-hydrodynamic evolution (e.g., \citep{Li1998, Metzger2010, Lippuner2015, Perego2022}). Some models implement Monte Carlo radiative transfer \citep{Kasen2017, Wollaeger2018, Bulla2019, Korobkin2021, Collins2023, Shingles2023}. Nuclear heating rates are typically derived from fits to nucleosynthesis calculations \citep{Wu2022, Rosswog2024, Ricigliano2024}.
Recent works simulate the radiation-hydrodynamics of the ejecta, improving the accuracy of ejecta evolution \citep{Roberts2011, Rosswog2014, Montes2016, Wu2022, Just2023, Kawaguchi2023, Kawaguchi2024}. The use of nuclear networks (NNs) sheds light on nucleosynthesis yields from BNSMs \citep{Bovard2017, Ricigliano2024} and provides high-fidelity estimates of the heating rate \citep{Lippuner2015, Lippuner2017}. The coupling of hydrodynamics and in-situ NNs improves the fidelity of the abundances and light curves predictions \citep{Magistrelli2024, Magistrelli:2025xja}.

Studies have directly observed $^{56}$Ni production in BNSM winds \citep{Bernuzzi2025, Jacobi2025}, whose mixing can modify the shapes of supernova light curves \citep{Nakar2016, Utrobin2019, Goldberg2019, Utrobin2021}. Ray-by-ray simulations \citep{Magistrelli2024} have observed large radial gradients in the composition of kilonova ejecta, raising the question of how composition mixing smooths these gradients and modifies the resultant nucleosynthesis and kilonova light curves. In this work, we implement composition mixing into ray-by-ray simulations coupled with the online NN \texttt{SkyNet}.
In Sec. \ref{sec: method}, we introduce the initial NR profiles and describe the hydrodynamic treatment, nucleosynthesis calculations, and composition mixing scheme. In Sec. \ref{sec: results}, we show results on how mixing influences the evolution of essential nuclear parameters, abundance yields, and light curves. We summarize and conclude our findings in Sec. \ref{sec: conclusion}.

\section{Methods} \label{sec: method}

\subsection{Initial Profiles}\label{sec: profile}

\begin{figure}[htb]
    \centering
    \includegraphics[width=1.0\linewidth]{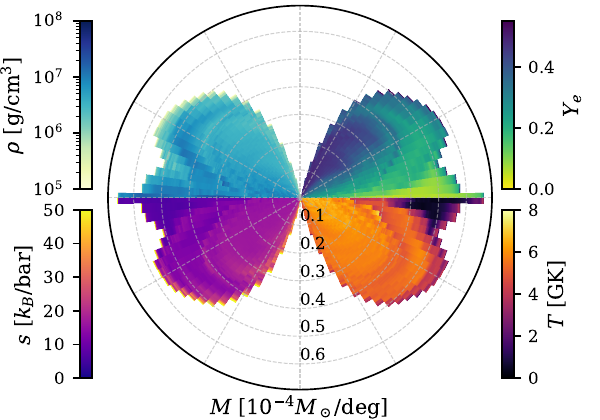}
    \caption{Initial conditions of the ejecta. The radial coordinate is the Lagrangian mass coordinate per unit polar angle, and the angular coordinate is the polar angle. The pole is in the vertical direction, and the equator is in the horizontal direction. From left to right and top to bottom, the quarter-spheres are density, electron fraction, entropy, and temperature, respectively. The rays in the polar regions are compressed in this plot since they have very low masses.}
    \label{fig: initial}
\end{figure}

Angular profiles are extracted from NR simulations of an equal-mass BNS merger with a gravitational mass of $1.35M_\odot$ per star \citep{Radice2023}. The merger was evolved by the NR \texttt{THC} code \citep{Radice2012, Radice2014a, Radice2014b}, with an M1 neutrino transport extension \citep{Radice2022, Zappa2023} and DD2 EoS \citep{Hempel2010, Typel2010}. The simulation ran at a spatial resolution ($\Delta x = 185\ \mathrm{m}$) to 100 ms post-merger, and we extract angular profiles at $R_\mathrm{ext} = 295\ \mathrm{km}$. The resolution of the NR simulation in the region covering the NSs before merger and the remnant is of $1.85 \times 10^4$ cm. The resolution in the
region where we extract ejecta data is of $5.9 \times 10^5$ cm. We collect unbound material according to the Bernoulli criterion \citep{Foucart2021}, resulting in a total mass of the ejecta of $7.38 \times 10^{-3} M_\odot$. Using the geodesic criterion to determine the dynamical ejecta \citep{Nedora2021, Combi2023, Ricigliano2024}, we identify a dynamical component of $2.69 \times 10^{-3}M_\odot$, and a disk wind component of $4.68 \times 10^{-3}M_\odot$.
The dynamical ejecta have a shock-heated component with high entropy, and a tidal component with low entropy.
The shock-heated component undergoes more intense neutrino captures, increasing its electron fraction to $Y_e > 0.1$. Additionally, the dense material in the equatorial plane shields the outer layers of the ejecta from neutrino irradiation, thereby maintaining low electron fractions. In contrast, at higher latitudes, the neutrino flux makes the ejecta more proton-rich \citep{Jacobi2024}. We show the distributions of density, electron fraction, entropy, and temperature in Fig. \ref{fig: initial}. The outer layers of the ejecta at the equator have low electron fraction and entropy.

We divide the ejecta into 51 azimuthally averaged angular profiles, with the properties of each profile dependent on the azimuthal angle integrated out. The profiles are mapped into the Lagrangian mass coordinates by placing layers on top of the most interior layer $R_\mathrm{ext}$. We scale masses of the profiles by a factor of $\lambda_\theta = 4\pi/\Delta\Omega$, where $\Delta\Omega = 2\pi\sin\theta\Delta\theta$ is the solid angle of the profile at the polar angle $\theta$ \citep{Magistrelli2024}, with all intensive properties unchanged. The profiles are further rearranged into 500 mass shells, with the finest grids at the inner and outer boundaries, to better resolve sharp density gradients and the rapid expansion. The initial radial location of the outer boundary varies between rays, ranging from $8.54 \times 10^{7}$ cm to $1.37 \times 10^{8}$ cm. The finest and coarsest radial grid spacings are $5.08 \times 10^{3}$ cm and $2.39 \times 10^{6}$ cm, respectively. Each profile is then evolved as an independent radiation-hydrodynamics spherically symmetric model. Eventually, we map the 1-D simulations back to the axisymmetric ejecta by multiplying all extensive properties by a factor of $1/\lambda_\theta$.

\subsection{Radiation-Hydrodynamics}

\begin{figure*}
    \centering
    \includegraphics[width = 0.99\linewidth]{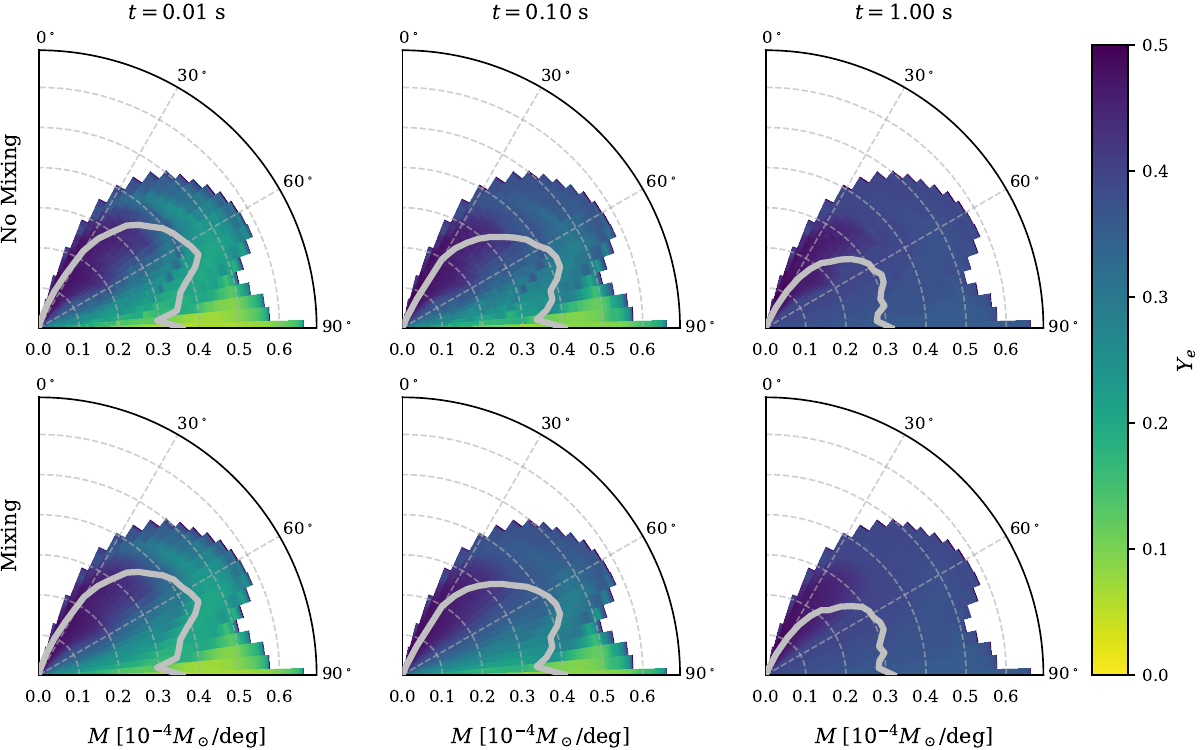}
    \caption{Evolution of the spatial distribution of the electron fraction $Y_e$. The radial coordinate is the Lagrangian mass coordinate per unit polar angle, and the angular coordinate is the polar angle. The top and bottom panels are the evolution of the no-mixing and mixing models, respectively. The three columns correspond to $t = 0.01,\ 0.1,\ \mathrm{and}\ 1\ \mathrm{s}$. The $Y_e$ distribution near the equator is robust under mixing, but in the rays with polar angles $\sim 60^\circ$, where $Y_e$ has a greater radial gradient, mixing smooths its distribution.
    The gray solid contour marks the surface of velocity of $0.1c$. As the ejecta expand and accelerate, the mass enclosed by this low-velocity surface decreases, indicating that a greater amount of ejecta has achieved velocities $>0.1c$.
    }
    \label{fig: Ye_map}
\end{figure*}

\begin{figure*}
    \centering
    \includegraphics[width = 0.99\linewidth]{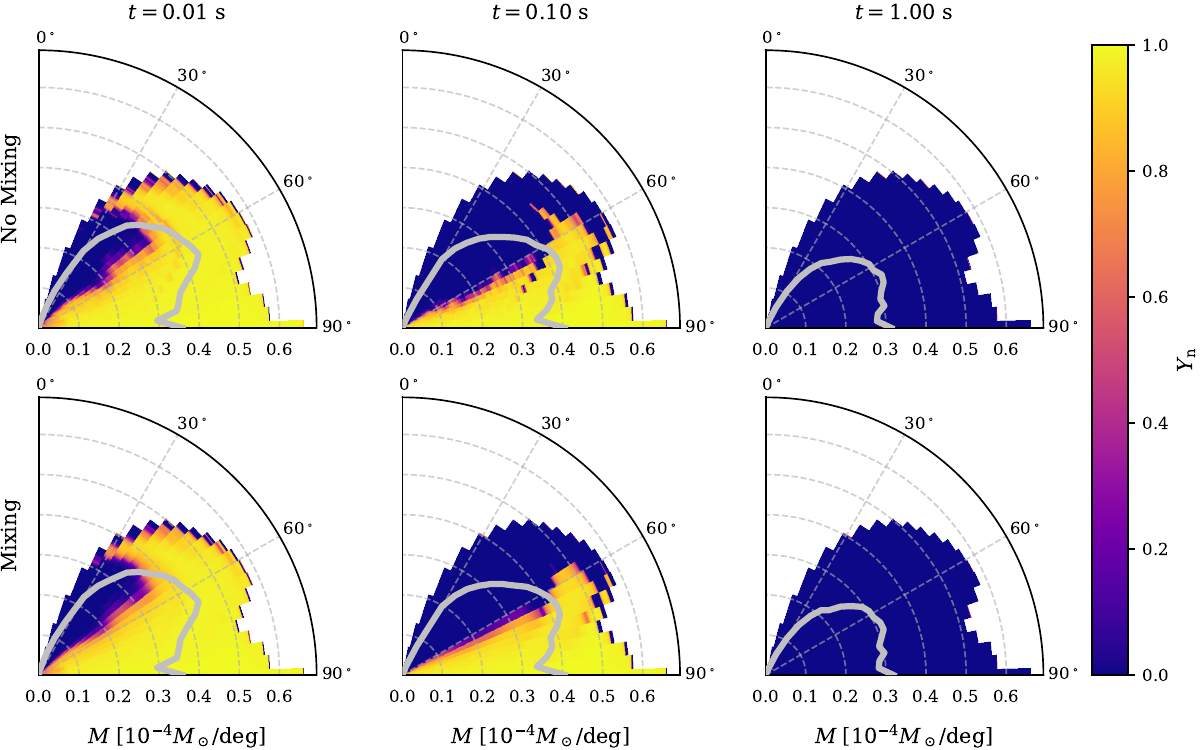}
    \caption{Evolution of the spatial distribution of the free-neutron abundance $Y_\mathrm{n}$. The coordinates, panels, and the gray contour follow Fig. \ref{fig: Ye_map}. Similar to $Y_e$, mixing influences $Y_\mathrm{n}$ the most at $\sim 60^\circ$, smoothing the jagged features in the no-mixing model. Free neutrons have already been exhausted at 1 s.}
    \label{fig: Yn_map}
\end{figure*}

For each profile, we evolve the radiation-hydrodynamics equations as described in \cite{Morozova2015, Wu2022, Magistrelli2024}. The mass conservation is:
\begin{equation}
    \frac{\partial r}{\partial m} = \frac{1}{4\pi r^2 \rho} \, .
\end{equation}
The energy conservation is:
\begin{equation}
    \frac{\partial \epsilon}{\partial t} = \frac{P}{\rho}\frac{\partial\ln \rho}{\partial t} - 4\pi r^2Q\frac{\partial v}{\partial m} - \frac{\partial L}{ \partial m} + \dot\epsilon_\mathrm{nuc}  \, .
\end{equation}
The momentum conservation equation is:
\begin{equation}
    \frac{\partial v}{\partial t} = -\frac{Gm}{r^2} - 4\pi r^2\frac{\partial P}{\partial m} - 4\pi \frac{\partial(r^2Q)}{\partial m}  \, .
\end{equation}
Here $r$ is the radius, $m$ is the mass coordinate, $\rho$ is the density, $\epsilon$ is the specific internal energy, $P$ is the pressure, $Q$ is the von Neumann–Richtmyer artificial viscosity \citep{Von-Neumann1950}, $v$ is the radial velocity, $L$ is the radiative luminosity, and $\dot\epsilon_\mathrm{nuc}$ is the local specific nuclear reaction power.

At low temperatures, the ejecta EoS follows the Paczy{\'n}ski EoS \citep{Paczynski1983}, which includes ideal ion gas, electron gas with ideal, non-relativistic degenerate and relativistic degenerate components, and ideal photon gas. At high temperatures, we use a tabulated Helmholtz EoS \citep{Timmes2000, Lippuner2017} implemented by \cite{Magistrelli2024}. We use the same analytic opacity as \cite{Wu2022} to calculate the luminosity $L$. The flux density is calculated as blackbody radiation from the photosphere and effective heating in the cells above the photosphere (see Eq. (14) of \cite{Wu2022}). Then, we convolve the flux density with filter functions to obtain light curves.

\subsection{Nuclear Network} \label{sec: NN}

To calculate the local nuclear reaction power $\dot\epsilon_\mathrm{nuc}$ and track the evolution of ejecta composition, we use the NN \texttt{SkyNet} \citep{Lippuner2017}, incorporated with the hydrodynamical evolution code by \cite{Magistrelli2024}. The NN is initialized with the temperature and density distribution, assuming nuclear statistical equilibrium (NSE) with initial abundances calculated as in \cite{Cowan2021, Perego2021}. We evolve the NN during each hydrodynamic timestep with its own sub-timesteps \citep{Magistrelli2024, Magistrelli:2025xja}. The temperature is constant in these sub-timesteps, while the density is log-interpolated between the start and end of the hydrodynamic timestep. We add the nuclear energy release to the internal energy and update the entropy change at the end of the NN evolution. The NN itself does not communicate the composition of different mass shells, so we must implement mixing between hydrodynamic timesteps.

\subsection{Composition Mixing Scheme}

In Lagrangian coordinates, we use mass fractions of isotopes rather than their number densities to maintain the constant masses of the shells. The changing rate of mass fractions is
\begin{equation}
    \frac{\partial X_i}{\partial t} = -\nabla \cdot J_i,
\end{equation}
where $X_i$ is the mass fraction of the $i$-th isotope and the mass fraction flux $J_i$ is
\begin{equation}
    J_i = -\lambda D\nabla X_i,
\end{equation}
in which $D$ is the mixing coefficient, and $\lambda$ is the flux limiter. We calculate the mixing coefficient by dimensional analysis, assuming that it depends on the density $\rho$, the sound speed $c_\mathrm{s}$, and a dimensionless parameter $\alpha$, and that mixing occurs within a lengthscale equal to the density scale height:
\begin{equation}
    \label{eq:D_def}
    D = \alpha c_\mathrm{s}\frac{\rho}{|\nabla\rho|}.
\end{equation}
We use $\alpha = 0.1$ in our simulations as we find it already leads to supersonic mixing. Mixing is more efficient when adjacent shells have similar densities, leading to smaller gradients in the denominator. We implement the flux limiter $\lambda$ \citep{Bersten2011} to avoid unrealistically large mixing speeds, limiting it to be within the subsonic regime:
\begin{equation}
    \lambda = \frac{6+3R}{6+3R+R^2},
\end{equation}
where
\begin{equation}
    R = \frac{w}{w_\mathrm{m}},
\end{equation}
in which $w = \max \left(D\frac{|\nabla X_i|}{X_i} \right)$ is the maximum mixing speed of all isotopes, and $w_\mathrm{m} = \frac{1}{3}c_\mathrm{s}$ is the upper limit of the speed for isotropic mixing.

Since the mixing equation is stiff, we numerically integrate it using the implicit Euler method \citep{Butcher2008}. The change of mass fractions from the $k$-th to the $(k+1)$-th timstep at the $j$-th mass shell is:
\begin{equation}
    \frac{X_{i,j}^{k+1} - X_{i,j}^k}{\Delta t^k} = \nabla \cdot (\lambda_j^{k+1} D_j^{k+1}\nabla X_{i,j}^{k+1}).
\end{equation}
Rather than explicitly calculating the flux of the current timestep, the implicit Euler method calculates the flux based on the mass fractions of the next timestep. This method maintains stable numerical evolution without requiring a smaller timestep than the hydrodynamic timestep. The gradient $\nabla X_{i,j}^{k+1}$ is evaluated at cell faces using the mass fractions at the two adjacent cell centers. The face fluxes are then used to compute the divergence $\nabla \cdot (\lambda_j^{k+1} D_j^{k+1}\nabla X_{i,j}^{k+1})$ at each cell center.

Although mixing itself is not computationally expensive, it slows down NN evolution because it forces the NN to adopt smaller time steps after composition changes. Thus, we turn off the mixing after $100\ \mathrm{s}$ of simulations, when the $r$-process reactions have stopped, to maintain computational efficiency. We first test mixing speed of $\alpha = 0.01,\ 0.1\ \mathrm{and}\ 1$, and a constant mixing coefficient. We observe that $\alpha = 0.1$ already makes the flux limiter active over most of the grid. Thus, we choose $\alpha = 0.1$ as a conservative upper limit on the effect of mixing for further exploration. In App. \ref{app: step_Ye} we introduce validation of our mixing scheme on an analytic electron fraction profile. We test turning off mixing using this profile with sharp $Y_e$ gradients, of which the nucleosynthesis yields and light curves are the same whether mixing is on or off after $100\ \mathrm{s}$.

\section{Results} \label{sec: results}

\begin{figure*}
    \centering
    \includegraphics[width = 1\linewidth]{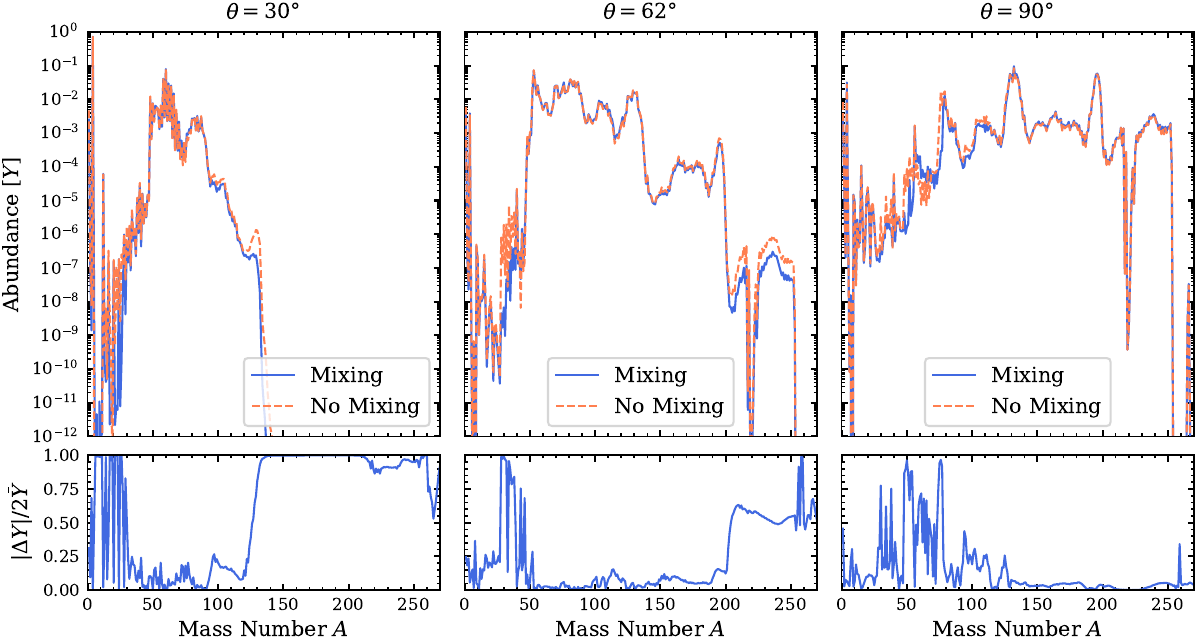}
    \caption{Final abundances of rays at polar angle $\theta = 30^\circ,\ 62^\circ,\ \mathrm{and}\ 90^\circ$ with reference to the mass number $A$. The relative residual in the bottom panel is defined as the absolute abundance difference divided by twice the average abundance of the mixing and no-mixing models. Mixing makes noticeable differences in the low-abundance elements of each ray, but the peaks are consistent.}
    \label{fig: Y_angle}
\end{figure*}

\begin{figure}[hbt]
    \centering
    \includegraphics[width = 1\columnwidth]{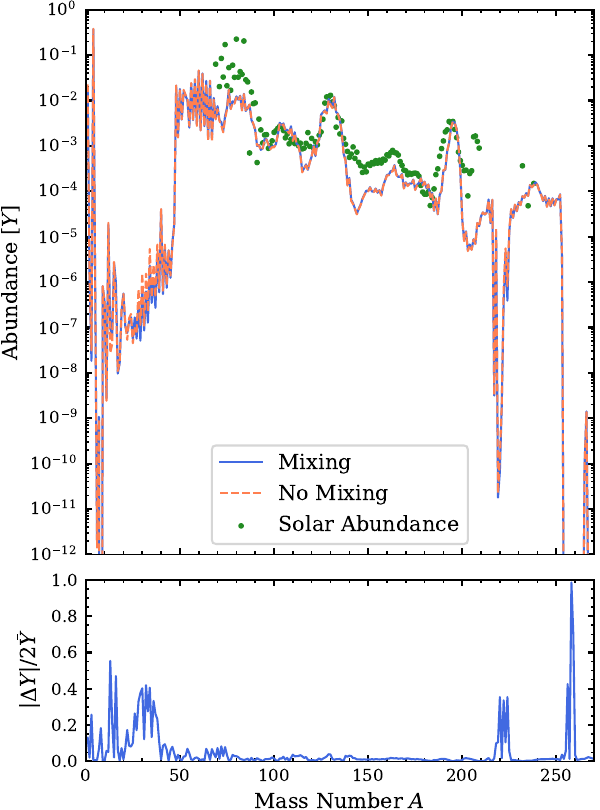}
    \caption{The combined global abundances with solar $r$-process residual abundances from \cite{Prantzos2020}, normalized at $A=195$. The axes are defined the same as in Fig. \ref{fig: Y_angle}. The abundances of the mixing and no-mixing models are consistent except for some light elements with $A<50$. The global yield is dominated by the angular rays where specific elements are produced most efficiently, which effectively dilutes the minor abundance differences found in other rays. The abundances of the second- and third-peak elements are comparable to the solar $r$-process abundances, but the first-peak and rare-earth elements are underproduced.}
    \label{fig: Y_merge}
\end{figure}

\subsection{Evolution under Mixing}\label{sec: evolution}

Initially, the angular rays near the equator have very low electron fractions $Y_e \lesssim 0.1$. At intermediate polar angles ($\theta \sim 45^\circ$), the outer layers could retain low electron fractions $Y_e \lesssim 0.3$, while the inner layers reach $Y_e \gtrsim 0.4$. The low $Y_e$ in these regions leads to a high abundance of free neutrons, $Y_\mathrm{n, 0} \gtrsim 0.8$, essential for $r$-process nucleosynthesis in low entropy conditions. Fig. \ref{fig: Ye_map} and \ref{fig: Yn_map} show the evolution of $Y_e$ and $Y_\mathrm{n}$, respectively, with and without mixing in all mass shells. The distributions of $Y_e$ and $Y_\mathrm{n}$ with mixing are generally consistent with no mixing, especially in the neutron-rich mass shells. The effects of mixing are more visible at the sharp interface around $\theta \sim 60^\circ$, where the sharp compositional gradients present in the no-mixing models (top panels) of Fig. \ref{fig: Yn_map} are smoothed out in the mixing model (bottom panels). However, mixing does not significantly change the
distribution of free neutrons near the equator, which is the primary site for strong $r$-process nucleosynthesis.

\subsection{Abundance Yields}

The production of $r$-process heavy elements follows the angular distribution of the electron fraction $Y_e$ and the free neutron abundance $Y_\mathrm{n}$. In Fig. \ref{fig: Y_angle}, we show the final ($t = 2\times10^6\ \mathrm{s}$) abundances of rays at polar angles of $30^\circ$, $62^\circ$, and $90^\circ$, respectively. In the bottom panel showing the relative residuals, the average abundance is defined as:
\begin{equation}
    \bar{Y} = \frac{1}{2}(Y_\mathrm{mixing} + Y_\mathrm{no\ mixing}),
\end{equation}
which normalizes the relative residual so that the maximum difference is 1. We can see an increase in heavy elements as more free neutron-rich shells are available. At $30^\circ$, with very few free neutrons, the ray fails to produce a large amount of second-peak ($A \sim 130$) and heavier elements. At $62^\circ$, the ejecta have initially abundant free neutrons, which are consumed very quickly and are almost gone at 0.1 s. Thus, its second and third-peak elements ($A \sim 195$) are not as plentiful as those of $90^\circ$. We do not see any significant difference in the abundances of $r$-process heavy elements between the mixing and no-mixing cases. Mixing is most influential on the elements that are marginally produced, thanks to their sensitivity to changes in free neutrons and their seeds.
However, the general $r$-process pattern is robust, as the global distributions of $Y_e$ and $Y_\mathrm{n}$ are stable under mixing (see Sec. \ref{sec: evolution}).

The mass-weighted final abundances are shown in Fig. \ref{fig: Y_merge}. At each mass number, the abundance is dominated by the rays that produce the associated species the most. However, the mixing induces differences only in subdominant isotopes (see Fig. \ref{fig: Y_angle}). As a result, the mixing effects are significantly diluted in the combined results. \cite{Prantzos2020} determined the $s$- and $r$-process components of the Solar system, with the absolute abundances measured by \cite{Lodders2009}. We show the solar $r$-process residual abundances as the scattered dots in Fig. \ref{fig: Y_merge}. They are normalized by the abundance at $A = 195$ in the no-mixing case. Results from our simulations are comparable with the solar $r$-process abundances at the second and third peaks in both mixing and no-mixing models. However, the third peak in the Solar system is broader than our simulation, whose shape is impacted by the $\beta$-decay rates \citep{Eichler2015, Kullmann2023}. Similar to \cite{Ricigliano2024} and \cite{Magistrelli2024, Magistrelli:2025xja}, with this scaling, we find the first-peak ($A \sim 82$) and the rare-earth elements with $135 \lesssim A \lesssim 165$ are underproduced. See App. \ref{app:abs_in_ang_rays} for further discussion on the angular dependencies of the abundances.

\begin{figure*}
    \centering
    \includegraphics[width = 0.99\linewidth]{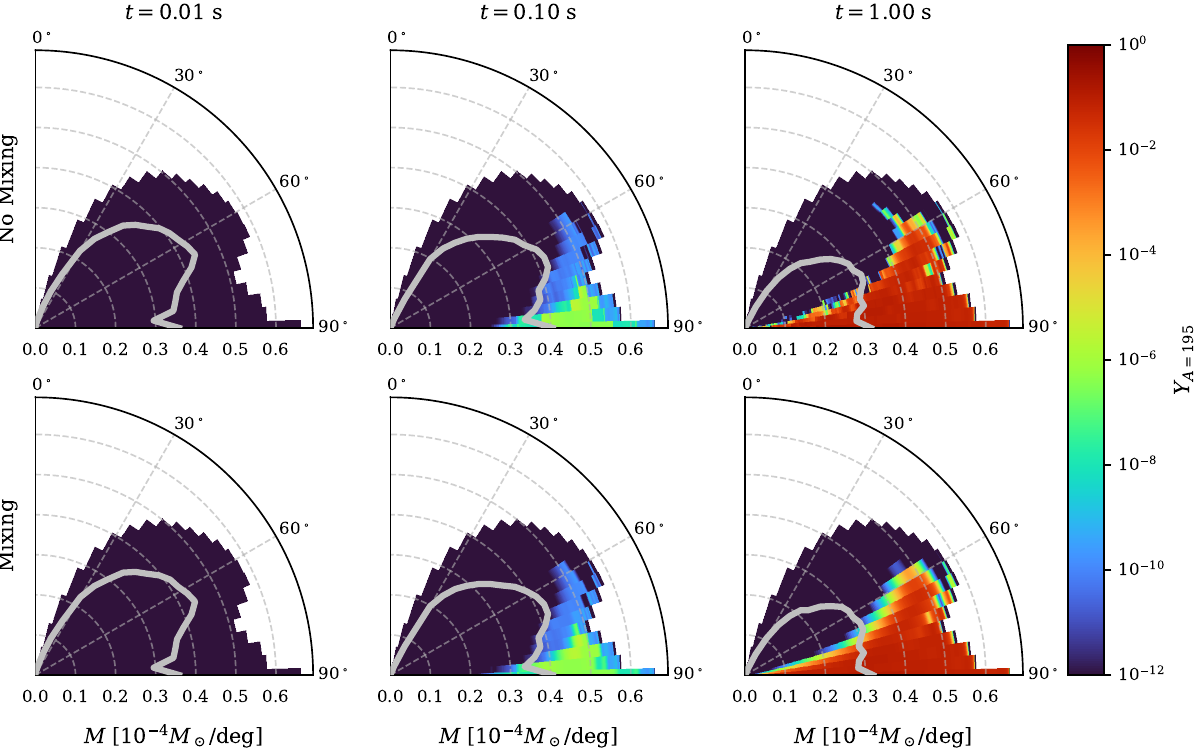}
    \caption{Evolution of the spatial distribution of abundances at $A=195$ (the third $r$-process peak) $Y_\mathrm{A=195}$. The coordinates, panels, and the gray contour follow Fig. \ref{fig: Ye_map}. The $Y_\mathrm{A=195}$ distribution at 1 s is generally consistent with the $Y_\mathrm{n}$ distribution at 0.1 s. The mixing makes noticeable differences only at the edges of high-abundance regions.}
    \label{fig: Y195_map}
\end{figure*}

\begin{figure*}
    \centering
    \includegraphics[width = 0.99\linewidth]{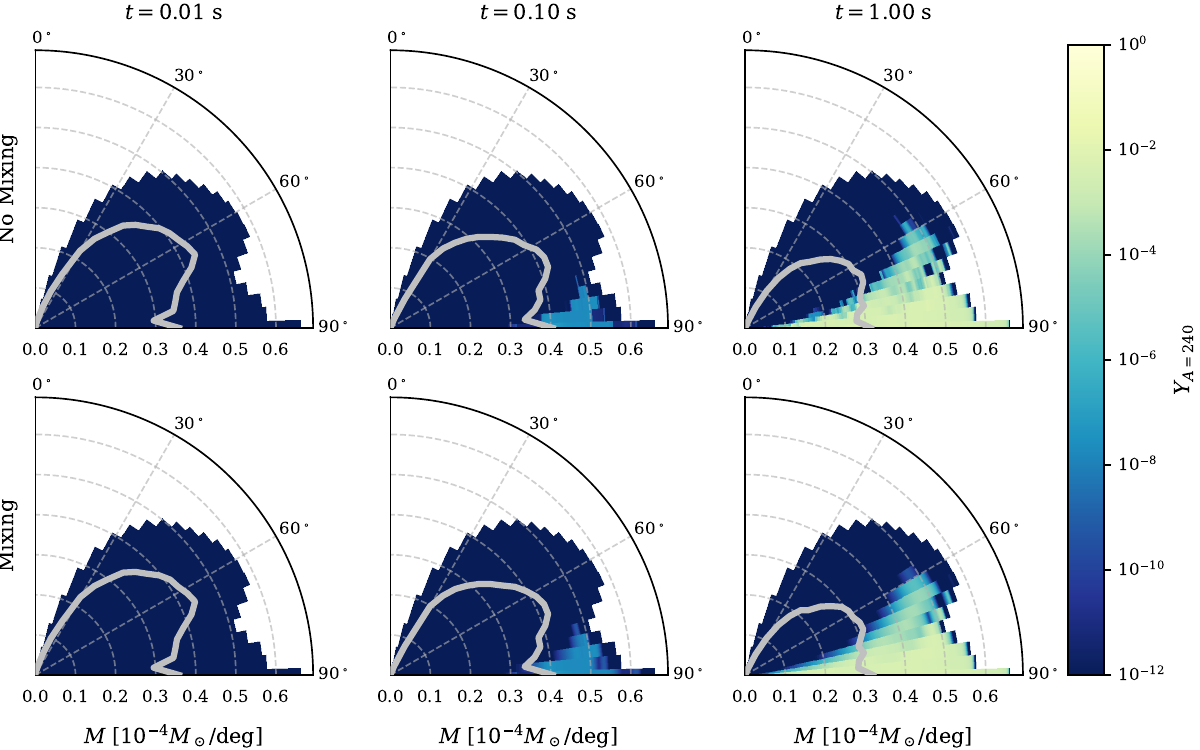}
    \caption{Evolution of the spatial distribution of abundances at $A=240$ (actinides) $Y_\mathrm{A=240}$. The coordinates, panels, and the gray contour follow Fig. \ref{fig: Ye_map}. Its distribution is similar to the $Y_\mathrm{A=195}$ distribution, also only noticeably influenced by mixing at the edges of high-abundance regions.}
    \label{fig: Y240_map}
\end{figure*}

In Fig. \ref{fig: Y195_map} and \ref{fig: Y240_map}, we show the evolution maps of abundances at mass numbers $A=195$ (the third peak) and $A=240$ (actinides), respectively. The $r$-process first produces heavy elements in the outer regions of the ejecta near the equator, where the ejecta have low entropy ($s \lesssim 10k_\mathrm{B}/\mathrm{baryon}$, \cite{Lippuner2017}) at the start of the simulation, and subsequently takes place in neutron-rich mass shells. The spatial distributions of these heavy elements at 1 s, when free neutrons have run out, match the spatial distributions of free neutrons at 0.1 s. As we have seen in the $Y_e$ and $Y_\mathrm{n}$ distributions in Fig. \ref{fig: Ye_map} and \ref{fig: Yn_map}, mixing influences the rays in the transition between rich and poor free neutron regions, where it smooths the fine-scale structures of the heavy elements, but does not alter the general spatial distribution and total global abundances.

\subsection{Kilonova Light Curves}

\begin{figure*}
    \centering
    \includegraphics[width = 1\linewidth]{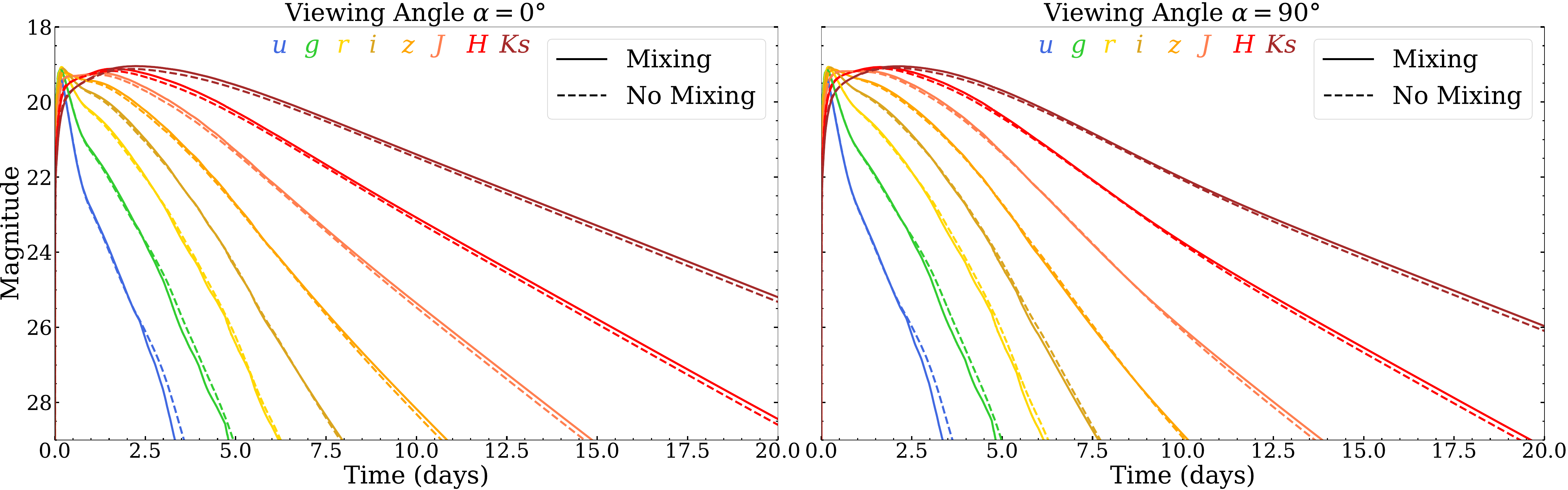}
    \caption{Light curves of the kilonova from the viewing angles $\alpha = 0^\circ$ (the left panel) and $90^\circ$ (the right panel) in the Gemini bands $u,\ g,\ r,\ i,\ z,\ J, H, \mathrm{and}\ K_s$, from optical to infrared. The kilonova has both blue and red components. The mixing model shows a slightly fainter blue tail and a slightly brighter red tail, but both are below the detection limit.}
    \label{fig: lc}
\end{figure*}

In Fig. \ref{fig: lc}, we show the light curves of the kilonova in the Gemini bands at viewing angles $\alpha = 0^\circ$ (pole) and $90^\circ$ (equator) and a distance of 40 Mpc. The filter functions are from the SVO Filter Profile Service \citep{Rodrigo2012, Rodrigo2020}. The kilonova has both blue and red components.
Globally, the mixing model deposits slightly more energy into the redder bands than the no-mixing model. Individual rays (especially the fast-decaying ones), however, exhibit a greater difference in their light curves. The rays with slow-decaying light curves dominate the combined light curves at late times, leading to a small difference between mixing and no-mixing models, due to the difference in the calculated nuclear powers. The differences in the light curves across all bands are well below the detectability of state-of-the-art telescopes, as they do not appear until the magnitudes have dropped significantly. They are also subdominant compared to other sources of uncertainty in the kilonova model (e.g., opacities, thermalization efficiency, etc.).

\section{Conclusion} \label{sec: conclusion}

In this work, we implemented composition mixing into ray-by-ray radiation-hydrodynamic kilonova simulations coupled with NN \citep{Wu2022, Magistrelli2024}. We ran simulations to study the influence of composition mixing on the evolution of critical nuclear parameters, element abundances, and kilonova light curves. We found that composition mixing is not a primary driver of uncertainty in kilonova simulations under the ray-by-ray assumption.

Mixing does not substantially alter the distributions of electron fractions ($Y_e$) and abundances of free neutrons ($Y_\mathrm{n}$) where they are relatively uniform (near the equatorial plane). The main difference appears around intermediate polar angles ${\sim} 60 ^\circ$, where $Y_e$ and $Y_\mathrm{n}$ change sharply in the radial and zenith directions. Since $Y_e$ and $Y_\mathrm{n}$ are generally robust under mixing in the rays, mixing only slightly perturbs the $r$-process, changing abundances of elements that are not dominant species. However, these differences are diluted in the global abundances, as the total yield is dominated by the angular rays where these elements are produced most efficiently.
The second and third $r$-process peaks are comparable to the solar $r$-process abundances \citep{Prantzos2020}. Still, the first-peak and rare-earth elements are underproduced, as seen in simulations by \cite{Magistrelli2024, Ricigliano2024, Magistrelli:2025xja}. The distributions of heavy $r$-process elements follow that of free neutrons at about 0.1 s. They exhibit similar behavior with $Y_\mathrm{n}$: mixing becomes influential only in regions with sharp abundance gradients, but it still fails to alter the global patterns. The kilonova light curves powered by $\beta-$decay of $r$-process elements are almost consistent with and without mixing. The mixing model only deposits slightly more energy into redder bands when the light curves have already decayed significantly.

Our work only considers radial mixing effects. For future work, we expect to explore mixing in full 3-D simulations of the BNSM ejecta over longer timescales, which might have more effects on nucleosynthesis and light curves. Additionally, a more accurate opacity scheme, such as \cite{Magistrelli:2025xja}, is needed, as the current analytic opacity relies solely on the initial $Y_e$, which may not reflect the realistic lanthanide and actinide opacities essential to kilonova spectra and light curves.

\begin{acknowledgments}
We are happy to thank Falk Herwig for the discussions that motivated this work.
RZ and DR were supported by the National Science Foundation under Grants
AST-2108467. DR also acknowledges support from the Sloan Foundation, from the
Department of Energy, Office of Science, Division of Nuclear Physics under Award
Numbers DE-SC0021177 and DE-SC0024388, from the National Science Foundation
under Grants No. PHY-2020275, PHY-2116686, PHY-2407681, and PHY-2512802.
FM acknowledges support from the Deutsche Forschungsgemeinschaft
(DFG) under Grant No. 406116891 within the Research Training Group
RTG 2522/1.
SB acknowledges funding from the EU Horizon under ERC Consolidator Grant,
no. InspiReM-101043372 and from the Deutsche Forschungsgemeinschaft, DFG,
project MEMI number BE 6301/2-1.
AP is supported by the European Union under NextGenerationEU, PRIN 2022
Project No.~2022KX2Z3B.
This research used resources of the National Energy Research Scientific Computing Center (NERSC), a Department of Energy User Facility using NERSC award
ERCAP0031370.
\end{acknowledgments}

\appendix


\section{Abundances in the Angular Rays}
\label{app:abs_in_ang_rays}

\begin{figure}[htb]
    \centering
    \includegraphics[width = 1\linewidth]{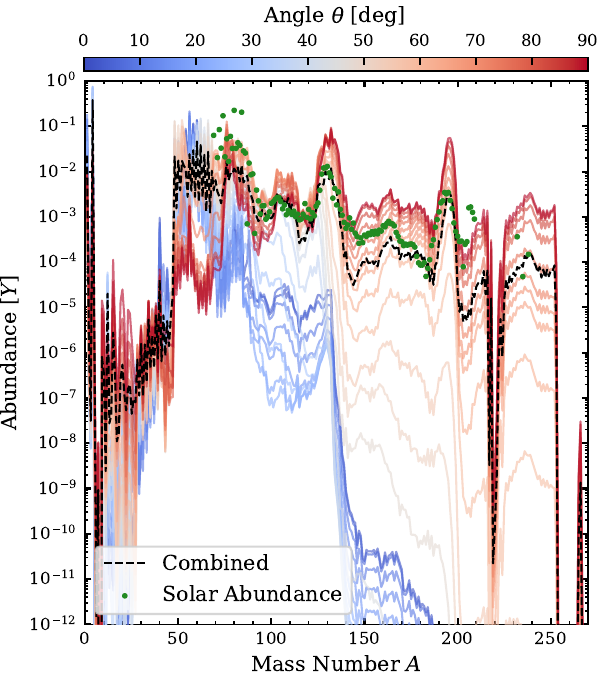}
    \caption{Abundances of angular rays with reference to the mass number $A$, with the combined global abundances of the no-mixing model and solar $r-$process abundances from \cite{Prantzos2020}. The solar $r-$process abundances are best matched by the rays at ${\sim} 70^\circ$, where they produce more first-peak and rare-earth elements, but do not produce an excess of second- and third-peak elements.}
    \label{fig: Y_angle_cmap}
\end{figure}

The relative abundances of elements could change dramatically with the viewing angle and influence the explanation of the abundances measured in the Solar system \citep[see e.g.][]{Chiesa2024}. In Fig. \ref{fig: Y_angle_cmap}, we show the abundances in the angular rays in the no-mixing case and the solar $r-$process abundances from \cite{Prantzos2020}. Abundances of the elements of the three $r-$process peaks are larger for rays closer to the equatorial plane. In the ray at $90^\circ$, the first-peak elements are underproduced compared with the rays at $\sim 70^\circ$, since they have larger neutron capture rates and are rapidly converted into heavier elements along very extended $r$-process paths. The solar abundances are more consistent with a mixture of the rays near the equatorial plane rather than the whole ejecta. The uncertainties in the nuclear reaction rates could also change the abundances of rare earths and the third peak \citep{Eichler2015, Kullmann2023}.

\section{Analytic Electron Fraction Profile}\label{app: step_Ye}

\begin{figure}[htb]
    \centering
    \includegraphics[width = 1\linewidth]{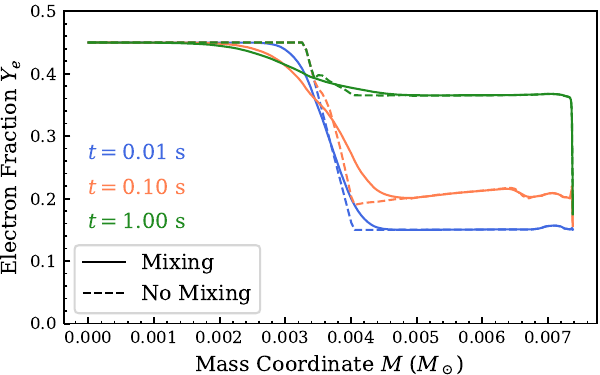}
    \caption{Evolution of the analytic $Y_e$ profile. The solid and dashed lines represent the mixing and no-mixing models, respectively. Blue, orange, and green correspond to the distributions at $t = 0.01,\ 0.1,\ \mathrm{and}\ 1\ \mathrm{s}$. Mixing smooths the sharp gradients and structures rising in the no-mixing model.}
    \label{fig: Ye_single}
\end{figure}

\begin{figure}[htb]
    \centering
    \includegraphics[width = 1\linewidth]{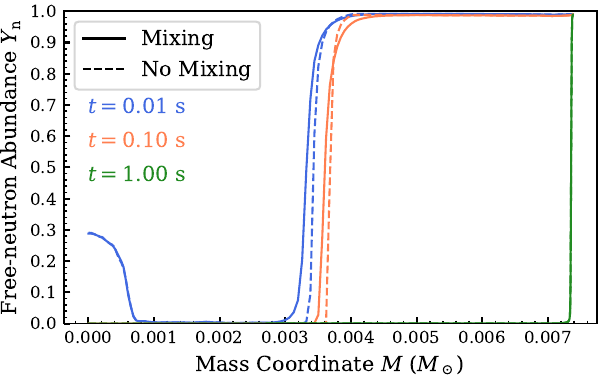}
    \caption{Evolution of free-neutron abundance $Y_\mathrm{n}$ in the case of the analytic $Y_e$ profile. The lines, shapes, and colors follow Fig. \ref{fig: Ye_single}. $Y_\mathrm{n}$ has a smaller gradient at the center of the mass coordinate under mixing.}
    \label{fig: Yn_single}
\end{figure}

\begin{figure}[htb]
    \centering
    \includegraphics[width = 1\linewidth]{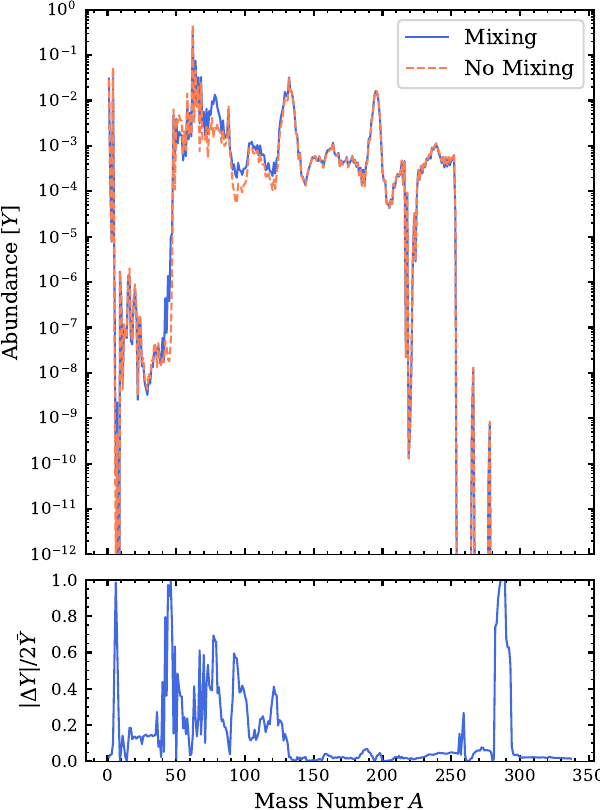}
    \caption{Final abundances of the analytic $Y_e$ profile. The coordinates are defined the same as in Fig. \ref{fig: Y_angle}. Mixing strongly changes the abundances of elements before the second $r-$process peak, but leaves heavier elements unaffected.}
    \label{fig: Y_single}
\end{figure}

We create an analytic profile for the electron fraction $Y_e$ to verify that our mixing scheme works as expected. This profile is an extreme case that is unrealistic in BNSM simulations.  We artificially redefine the inner $Y_e$ to 0.45 and the outer $Y_e$ to 0.15, interpolating linearly between these two values over the mass range $0.45\leq M/M_\mathrm{total} \leq 0.55$. This prescription provides a sharp transition in this range and improves the resolution of mixing effects. We directly take all other physical properties from the NR profile (see Sec. \ref{sec: profile}), and then initialize the NN as described in Sec. \ref{sec: NN}.

In Fig. \ref{fig: Ye_single} and \ref{fig: Yn_single}, we show the evolution of the electron fraction $Y_e$ and the free-neutron abundance $Y_\mathrm{n}$, respectively. Mixing smooths the initially sharp boundary between the high- and low-$Y_e$ regions (neutron-poor and neutron-rich regions). Additionally, it removes the complex structures in $Y_e$ that arise at the boundary in the no-mixing model. We compare the abundances of mixing and no-mixing models in Fig. \ref{fig: Y_single}. Mixing changes the abundances of elements before the second $r-$process peak, leaving heavier elements unaffected.

Overall, mixing is efficient in regions with strong abundance gradients but fails to significantly influence the more uniform neutron-rich regions. Thus, abundances of the heaviest elements stay consistent in the final nucleosynthesis output.

\bibliographystyle{apsrev4-2}
\bibliography{Zhai}

\end{document}